# Evidences for the exciton gas phase and its condensation in monolayer 1T-ZrTe$_2$

Yekai Song[1,2,†], Chunjing Jia[3,4,†], Hongyu Xiong[3,4,5], Binbin Wang[6], Zhicheng Jiang[1], Kui Huang[6], Jinwoong Hwang[4,7,8], Zhuojun Li[1,2], Choongyu Hwang[8], Zhongkai Liu[6], Dawei Shen[1], Jonathan A. Sobota[3,4], Patrick Kirchmann[3,4], Jiamin Xue[6], Thomas P. Devereaux[3,4], Sung-Kwan Mo[7*], Zhi-Xun Shen[3,4*], Shujie Tang[1,2*]

[1]State Key Laboratory of Functional Materials for Informatics, Shanghai Institute of Microsystem and Information Technology, Chinese Academy of Sciences, Shanghai 200050, China.

[2]2020 X-Lab, Shanghai Institute of Microsystem and Information Technology, Chinese Academy of Sciences, Shanghai 200050, China.

[3]Stanford Institute for Materials and Energy Sciences, SLAC National Accelerator Laboratory, 2575 Sand Hill Road, Menlo Park, California 94025, USA.

[4]Geballe Laboratory for Advanced Materials, Departments of Physics and Applied Physics, Stanford University, Stanford, California 94305, USA.

[5]Key Laboratory for Power Machinery and Engineering of MOE, School of Mechanical Engineering, Shanghai Jiao Tong University, Shanghai 200240, PR China.

[6]School of Physical Science and Technology, ShanghaiTech University, Shanghai 201210, China.

[7]Advanced Light Source, Lawrence Berkeley National Laboratory, Berkeley, California 94720, USA.

[8]Department of Physics, Pusan National University, Busan 46241, Korea.

*Corresponding authors. Email: SKMo@lbl.gov, zxshen@stanford.edu, tangsj@mail.sim.ac.cn

**The excitonic insulator (EI) is a Bose-Einstein condensation (BEC) of excitons bound by electron-hole interaction in a solid, which could support high-temperature BEC transition[1-3]. The material realization of EI has been elusive, which is further challenged by the difficulty of distinguishing it from a conventional charge density wave (CDW) state. In the BEC limit, the pre-condensation exciton gas phase[4] is a hallmark to distinguish EI from conventional CDW, yet direct experimental evidence has been lacking[5]. Here we report a distinct correlated phase beyond the 2×2 CDW ground state emerging in epitaxially grown monolayer 1T-$ZrTe_2$ and its investigation by angle-resolved photoemission spectroscopy (ARPES) and scanning tunneling microscopy (STM). The results show novel band- and energy-dependent folding behavior in a two-step process, evidenced by an exciton gas phase prior to its condensation into the final CDW state. The excellent agreement between experiments and theoretical predictions on the recovery of the pristine band structure by carrier-density-dependent suppression of the CDW state further corroborates the monolayer 1T-$ZrTe_2$ as an EI. Our findings provide a versatile two-dimensional platform that allows tuning of the excitonic effect.**

The macroscopic quantum phenomena such as superconductivity and BEC are crucial and attractive in basic research and potential technological applications. The EI was originally proposed as an analog of Bardeen-Cooper-Schrieffer (BCS) condensate in a weak-coupling limt[1-3]. It may support novel transport properties[6,7] corresponding to the superfluidity in superconductors. With varying coupling strength, a wealth of exotic phenomena could stem from EIs[8,9], such as correlated Chern insulator and fractionalized Hall effect[10]. However, the unequivocal experimental evidence for the realization of an EI state is still lacking. Unlike singlet-paired superconductors, the electron-hole pair in exciton does not carry a net charge, thus providing no easy observable physical quantity. Further complicating the issue, the EI state with finite-momentum exciton is characterized by a CDW state with periodic lattice distortion (PLD). Hardly any difference is present in the ground state electronic structure due to the same symmetry-breaking behavior in the electronically driven EI and a conventional CDW state from electron-phonon coupling (EPC).

Experimental efforts to identify the excitonic effect have been devoted to looking beyond the static electronic structure, such as the softening of the plasmon mode in BCS limit EI candidate[11] and ultra-fast melting of the CDW order upon laser pumping[12], or constructing van der Waals heterostructures and separating electron and hole layers to avoid the fast recombination[13,14]. Another distinct characteristic of excitonic condensate, in contrast to the conventional CDW state, is the existence of the exciton gas state in the BEC limit. However, due to the very limited candidate materials, the experimental observation of the exciton gas phase is still left to be explored[5].

Reduction of the dimensionality enhances the excitonic effect by suppressing the screening, as evidenced in the unprecedentedly strong excitonic effect of the monolayer transition metal dichalcogenides (TMDC)[15,16] and other low dimensional materials[17]. The semi-metallic 1T-ZrTe$_2$[18-21], a sister compound of EI candidate 1T-TiSe$_2$, hosts a CDW state when thinning down to its 2D limit[22,23]. Compared to 1T-TiSe$_2$, the absence of CDW in bulk indicates the suppressed influence of EPC and a potential advantage in investigating the excitonic instability stemming from the characteristic band structure that nurtures the EI[2].

Using molecular beam epitaxy (MBE), we have successfully grown high-quality monolayer 1T-ZrTe$_2$. In-situ ARPES and STM measurements confirmed the emergence of a 2×2 CDW ground state in low temperature (LT)[22,23], and revealed a strong spectral weight (SW) transfer and band folding. Using carrier doping to suppress the interaction, we uncovered the pristine electronic structure of the system before excitonic interactions modified it. This insight reveals a hidden phase distinct from the ultimate EI with CDW formation. This two-step process is further confirmed by the melting of the CDW in higher temperature that reveal band- and energy-dependent relaxation of the band folding, exposing a temperature regime well above the CDW transition temperature ($T_c$) but with exciton bands fully developed. Such observations decisively deviate from the conventional CDW systems, but naturally occur in the preformed exciton gas scenario[4], which is rooted in the fundamental aspect that the PLD is a secondary effect in the excitonic condensation.

Figure 1a represents the crystal structure of the monolayer 1T-ZrTe$_2$, where two Te layers sandwich a Zr layer with octahedral coordination. Sub-monolayer 1T-ZrTe$_2$ were grown on bilayer graphene (BLG) terminated SiC substrate, which is known to induce the least amount of interaction between the MBE grown film and substrate[24]. Figs. 1b and c show the reflection high-energy electron diffraction (RHEED) and low energy electron diffraction (LEED) images of the film, respectively, showing the hexagonal symmetry of the 2D crystal with the in-plane orientation aligned with the BLG substrate. Using the lattice constant of BLG ($a$ = 2.46 Å) as a reference, the lattice constant of the film is estimated to be 4.0 ± 0.2 Å, consistent with the value of bulk 1T-ZrTe$_2$. The large-scale STM image at room temperature in Fig. 1d shows the typical morphology of monolayer 1T-ZrTe$_2$ domains, with the typical lateral size of ~ 50 nm.

At $T$ = 4.5 K, a 2×2 superlattice is observed by an atomically resolved STM image (Fig. 1e), and its 2D Fast Fourier transform (FFT) image (the insert in Fig. 1e), indicates the occurrence of CDW order at LT. ARPES spectra (Fig. 1f) taken from the monolayer 1T-ZrTe$_2$ along the Γ-M direction further confirm the existence of the CDW state. For the 2×2 CDW state, the first and second valence bands below Fermi energy ($E_F$) at the Γ point are folded into the M point. The SW transfer is distinctive in the system: the SW of the low energy valence band at the Γ point is nearly depleted around its maximum, with most of the SW transferred to the folded band at the M point, while the SW of the conduction band at the M point is kept nearly intact close to $E_F$. Notably, such observations were also made in bulk 1T-TiSe$_2$, attributed to the formation of the EI[25,26], which provides a corresponding reference to consider monolayer 1T-ZrTe$_2$ as an EI candidate material.

The temperature evolution of the CDW state in monolayer 1T-ZrTe$_2$ was studied using STM and ARPES. Fig. 2a shows the melting of the CDW order for selected temperatures from $T$ = 4.5 K to $T$ = 300 K. With increasing temperature, the superlattice contrast in the real space gets blurry (77 K) and then hardly visible (116 K). The 2×2 superlattice peaks in the FFT get weaker and diffused accordingly. Above some critical temperature, the superlattice peaks disappear, and only the Bragg peaks can be observed (a typical result at 300 K is shown in Fig. 2a). The melting of the CDW order is also characterized in ARPES by the bandgap (defined by the energy difference between the conduction and folded valence bands) shrinking and the disappearance of the folded second valence bands at the M point (Figs. 2b & 2c). As temperature increases, the folded first valence band gradually shifts to lower binding energy and merges with the conduction band. The folded second valence peak also slightly moves towards $E_F$, and its SW (shaded area) in Fig. 2c gradually fades away up to $T$ = 150 K. Note the presence of another feature from -0.2 eV to -0.8 eV that persists in all temperatures. This is likely related to the existence of the impurity phase or defect states[27], which is not relevant to the current discussion.

The temperature evolution of bandgap size, the intensity of the folded second valence band, and the intensity of the normalized 2×2 superlattice peak are plotted in Fig. 2d. These three observables are indicators of the CDW order, showing the identical temperature evolution. The $T_c$ in the monolayer 1T-ZrTe$_2$ is thus determined to be about 130 ± 20 K. A detailed look into the valence band at 300 K (Fig. S3 in the Supplementary Information), on the other hand, reveals a surprise that the top of the first valence band is well below $E_F$. Here, a strong SW transfer of the first valence band, i.e., the depletion of SW at the Γ point, persists well above $T_c$ as in the 300 K data of Fig. 2b, which does not follow the same trend as the other three indicators. Such asynchronous behavior of the first valence band implies the deviation from the conventional CDW picture where band modification and CDW formation goes hand in hand.

To investigate the nature of the asynchronous band folding, we use carrier doping to restore the screening in monolayer[18] and suppress the many-body interaction[24,28]. Upon surface K (potassium)-doping[29], the CDW order decreases drastically and eventually dissolves completely (Fig. 3e). Compared to the CDW state, a prominent valence band modification, from a flat band top to a nearly linear dispersion, significantly shrinks the bandgap from ~ 70 meV to -100 meV, with a concomitant recovery of SW at the Γ point. To rule out the potential band distortion that K-doping may introduce[30], we also employed photo-excitation with laser pulses to inject the carrier into the system to mimic the process of screening building up. The band dispersions before (t = -5 ps) and after (t = 100 fs) pumping at 12 K are plotted in Fig. 3g, exhibiting a transformation from a flattened valence band reshaped to a nearly linear dispersion upon pumping at the Γ point, which is identical to the K-doped case. The carrier injection suppresses the many-body effect and leads to an interaction-suppressed state, which is corroborated by reproducing the main electronic features of the noninteracting band structure from DFT calculations (Fig. S4 in the Supplementary Information).

The direct experimental confirmation of the noninteracting state is especially important, which enables the disclosure of the existence of a correlated state prior to the CDW state. Comparing to the well screened state, the first valence band dispersion of the state at 300 K flattened prominently, which leads to a global bandgap opening and an accompanied strong SW transfer (Figs. 3i and 3j). Since the temperature is far above the CDW $T_c$, the gap opening is less relevant to the CDW formation. In contrast, both characteristics constitute evidence for the existence of preformed exciton gas in an EI candidate material. First, the preformed exciton contributes a spectral gap, like the cooper pair does in the superconductors.[4] In addition, the exciton dispersion in the ARPES is under the conduction band minimum, whereas the SW of valence band maximum is transferred. Thus, the experimental observation evidence the correlated state as a preformed exciton gas phase beyond the CDW state.

As summarized in a schematic band diagram (Fig. 4a), the CDW formation is a two-step process, and the second valence band folding occurs only in the CDW state, indicating a different driving force from the first valence band folding. In the following, we show it is a consequence of the appearance of PLD driven by the exciton gas condensation, by excluding other possible alternative explanation. First, we investigate the different roles that played by PLD and excitonic effect in driving the band folding by the model calculations based on the microscopic Hamiltonians obtained from first-principles calculations, as shown in Fig. 4b (see Methods). For the PLD case, the folded SW from the first two valence bands with similar orbital components are both prominent[31] (Fig. 4b left and 4c left). While for the exciton gas state, the excitonic interaction is much more energy- and band-dependent, making the band folding concentrated on the top of the first valence band and the bottom of the conduction band (Fig. 4b right and 4c right). One alternative explanation of the two-step CDW formation could be the strong fluctuation in the 2D CDW system, which leads to short-range CDW formation above $T_c$. However, fluctuation itself is incapable of containing the asynchronous band folding behavior. The folding behavior of different valence bands should always be consistent because of the same lattice distortion origin in conventional CDW, which is contrary to the experimental observations. But in the BEC type of the exciton condensation, the two-step CDW formation and the asynchronous band folding behavior occur naturally. Above the BEC transition temperature $T_c$, the excitonic effect hybridizes and renormalizes the valence and conduction bands[4], and the preformed exciton is characterized by a folded valence band as seen in photoemisson[5]. Below $T_c$, the exciton condensation drives a long-range CDW order formation, with PLD as secondary effect[26]. On top of the purely electronically driven first valence folding, a second valence band folding emerges upon the CDW formation.

In addition to the distinct temperature dependence of the two valence band foldings, an identical two-step process with the asynchronous band folding is also found by changing the doping amount, as indicated in the plot of the SWs of the first valence band top at the Γ point and the folded second valence band top at the M point as a function of carrier density amount in Fig. 4d. In which, the SW transfer of the second valence band, the indicator of the long-range CDW formation, is found to be very sensitive to the doping level. A critical doping amount of 2% electron per unit cell could suppress the SW transfer, which is also consistent with the expectation of the excitonic instability induced CDW[29].

In summary, our experimental observation of the exciton gas phase in the monolayer 1T-ZrTe$_2$ establishes it as a promising EI candidate material with a highly tunable electronic structure. Moreover, its layered nature is advantageous in constructing van der Waals heterostructures, which is promising in exploiting excitonic physics, such as realizing spin supercurrent by embracing magnetism[32].

## Methods

**Film growth and STM measurements.** The monolayer 1T-ZrTe$_2$ films were grown by MBE on bilayer graphene (BLG) epitaxially grown on 4H-SiC. The base pressure of the system was ~ 5×10$^{-10}$ mbar. Zr (99.95%) and Te (99.999%) were evaporated from an electron beam evaporator and thermal cracker cell, respectively. The substrate temperature was held at 330 °C during growth. The growth process was monitored by RHEED. After growth, the samples were transferred under vacuum to STM (VT, Omicron) chamber with a base pressure on ~ 3×10$^{-10}$ mbar for in situ characterization. Some samples were transferred through a vacuum suitcase to STM (LT, Omicron) for low temperature characterization. STM images were acquired at all temperatures using W-tips.

**ARPES measurements.** In-situ ARPES measurements were performed at Beamline 10.0.1, Advanced Light Source, Lawrence Berkeley National Laboratory. ARPES data were acquired by Scienta R4000 electron analyzer. The system energy resolution and the angular resolution were set to 12 meV and 0.2°. The potassium was evaporated from a SAES Getters alkali metal dispenser to surface dope the thin film samples at the temperature of 10 K. ARPES measurements were also performed at the 03U beamline of the Shanghai Synchrotron Radiation Facility (SSRF) equipped with Scienta-Omicron DA30 electron analyzer. The angular and the energy resolutions were set to 0.2° and 8 ~ 20 meV (dependent on the selected probing photon energy).

**trARPES measurements.** The grown films were transferred into the trARPES load-lock (pressure 1×10$^{-9}$ Torr) through a vacuum suitcase with a base pressure of 3×10$^{-10}$ Torr. The trARPES measurements were based on a Ti: Sapphire regenerative amplifier operating at a repetition rate of 300 kHz. We used 1.5 eV linearly polarized IR pulse to excite the sample and used 6.0 eV UV pulse to probe the transient populations of the occupied and unoccupied band structure at a variable delay time. The overall time resolution of 70 ± 5 fs was extracted from cross-correlations of pump and probe pulses, and $t_0$ refers to both pulses overlapping in time. The beam profiles for IR and UV were 69×10² μm² and 23×30 μm², respectively. The photo-emitted electrons were collected by a Scienta R4000 analyzer in an ultrahigh vacuum with a base pressure less than 7×10⁻¹¹ Torr. The energy resolution was 40 meV. During the measurement, the sample temperature was maintained at 12 K.

**Theoretical calculation.** The theoretical result as shown in the left panel of Fig. 4b was calculated using the following method: Firstly, the structural relaxation was performed for the 2×2 supercell of the ZrTe$_2$ structure using density functional theory calculation with generalized gradient approximation (GGA) functional implemented in quantum espresso[35]. The relaxed structure showed the CDW pattern. Secondly, the effective microscopic Hamiltonian for the CDW supercell, including the Zr 3*d* and Te 2*p* orbitals, was obtained using the Wannier downfolding, implemented in Wannier90[36]. An energy shift of 1 eV has been implemented on the on-site energies of the Zr 3*d* orbitals, in order to make the bandgap of the simulations consistent with the experiment. Thirdly, the bandstructure was calculated for the obtained tight-binding model Hamiltonian, with the wavefunctions projected on the primitive Brillouin zone associated with the original unit cell. Red and blue weights represent the orbital content for Zr and Te respectively. The theoretical result as shown in the right panel of Fig. 4b was calculated using the following method: Firstly, the density functional theory calculation using quantum espresso[35] and the Wannier downfolding for the Zr 3*d* and Te 2*p* orbitals using Wannier90[36] were implemented for the unit cell of ZrTe$_2$, obtaining the effective microscopic Hamiltonian. An energy shift of 1eV has been applied to the on-site energies of the Zr 3$d$ orbitals to make consistent with the experimental bandstructure. Secondly, the bandstructure for the primitive Brillouin zone was obtained using the tight-binding model Hamiltonian. Thirdly, the excitonic interactions were added between the top of the first (or second) valence band and the bottom of the first conduction band with momentum q difference. The Hamiltonian could be written as:

$$H = H_{tight-binding} + \sum_{q=q1,q2,q3}\sum_{k} \Delta_1 c^+_{k,v1} c_{k-q,c1} + \Delta_2 c^+_{k,v2} c_{k-q,c1} + h.c.$$

in which $H_{tight-binding}$ is the tight-binding Hamiltonian obtained using Wannier90, $q1 = (0.0, -0.5), q2 = (0.5, 0.0), q3 = (-0.5, 0.5)$, $v1$ and $v2$ represent the first (top) and the second valence bands respectively, $c1$ represents the first conduction band, $\Delta_1 = 0.07eV$ and $\Delta_2 = 0.005eV$ in the calculation. The obtained bandstructure of this Hamiltonian was plotted on the right panel of Fig. 4B. Red and blue weights represent the orbital content for Zr and Te respectively.

## Acknowledgements

Research performed at ALS (thin film growth and ARPES) is supported by the Office of Basic Energy Sciences, US DOE under Contract No. DE-AC02-05CH11231. The work at Stanford University and SLAC National Accelerator Laboratory (thin film characterization, theory calculations) was supported by the Office of Basic Energy Sciences, US DOE under Contract No. DE-AC02-76SF00515. Part of this research used Beam line 03U of the Shanghai Synchrotron Radiation Facility, which is supported by $ME^2$ project under Contract No. 11227902 from National Natural Science Foundation of China. A portion of the computational work was performed using the resources of the National Energy Research Scientific Computing Center (NERSC) supported by the U.S. Department of Energy, Office of Science, under Contract No. DE-AC02-05CH11231. S. T. acknowledges the financial support from the National Natural Science Foundation of China (No. 11704395, No. 11974370), 'Strategic Priority Research Program (B)' of the Chinese Academy of Sciences (No. XDB04010600). D. S. acknowledges the support by National Natural Science Foundation of China (Grant No. U2032208). J. H. and C. H. acknowledge support from the NRF grant funded by the Korean government (MSIT) (No. 2020K1A3A7A09080369 and 2021R1A2C1004266).

## Author contributions

S. T., S.-K. M. and Z.-X. S. proposed and designed the research. Y. S. and S. T. performed the MBE growth. Y. S. and S. T. carried out the ARPES measurements and analyzed the ARPES data with help from Z. J., D. S., K. H., J. H., and S.-K. M.  Z. L. H. X., J. S. and P. K. carried out the trARPES measurements and analyzed the trARPES data. Y. S., B. W. and J. X. performed the STM measurements and analyzed STM data.

C. J. and T. P. D. carried out the density functional calculations and provided theoretical support. Y. S., S. T., S.-K. M. and Z.-X. S. wrote the manuscript with contributions and comments from all authors.

## Competing interests

The authors declare no competing financial interests.

# Reference

1.  Mott, N. F. The transition to the metallic state. *The Philosophical Magazine: A Journal of Theoretical Experimental and Applied Physics* **6**, 287-309 (1961).

2.  Jérome, D., Rice, T. M. & Kohn, W. Excitonic Insulator. *Phys. Rev.* **158**, 462-475 (1967).

3.  Kohn, W. Excitonic Phases. *Phys. Rev. Lett.* **19**, 439-442 (1967).

4.  Rustagi, A. & Kemper, A. F. Photoemission signature of excitons. *Phys. Rev. B* **97**, 235310 (2018).

5.  Fukutani, K. *et al.* Detecting photoelectrons from spontaneously formed excitons. *Nat. Phys.* **17**, 1024-1030 (2021).

6.  Eisenstein, J. P. & MacDonald, A. H. Bose-Einstein condensation of excitons in bilayer electron systems. *Nature* **432**, 691-694 (2004).

7.  Lozovik, Y. E. & Yudson, V. I. Feasibility of Superfluidity of Paired Spatially Separated Electrons and Holes - New Superconductivity Mechanism. *JETP Lett.* **22**, 274-276 (1975).

8.  Zenker, B., Ihle, D., Bronold, F. X. & Fehske, H. Slave-boson field fluctuation approach to the extended Falicov-Kimball model: Charge, orbital, and excitonic susceptibilities. *Phys. Rev. B* **83**, 235123 (2011).

9.  Kunes, J. Excitonic condensation in systems of strongly correlated electrons. *J. Phys.: Condens. Matter* **27**, 333201 (2015).

10. Hu, Y., Venderbos, J. W. & Kane, C. Fractional Excitonic Insulator. *Phys. Rev. Lett.* **121**, 126601 (2018).

11. Kogar, A. *et al.* Signatures of exciton condensation in a transition metal dichalcogenide. *Science* **358**, 1314-1317 (2017).

12. Rohwer, T. *et al.* Collapse of long-range charge order tracked by time-resolved photoemission at high momenta. *Nature* **471**, 490-493 (2011).

13. Wang, K. *et al.* Electrical control of charged carriers and excitons in atomically thin materials. *Nat. Nanotechnol.* **13**, 128-132 (2018).

14. Ma, L. *et al.* Strongly correlated excitonic insulator in atomic double layers. *Nature* **598**, 585-589 (2021).

15. Jia, Y. *et al.* Evidence for a monolayer excitonic insulator. *Nat. Phys.* **18**, 87-93 (2022).

16. Sun, B. *et al.* Evidence for equilibrium exciton condensation in monolayer $WTe_2$. *Nat. Phys.* **18**, 94-99 (2022).

17. Ma, J. *et al.* Multiple mobile excitons manifested as sidebands in quasi-one-dimensional metallic $TaSe_3$. *Nat. Mater.* 1-7 (2022).

18. Tsipas, P. *et al.* Massless Dirac Fermions in $ZrTe_2$ Semimetal Grown on InAs(111) by van der Waals Epitaxy. *Acs Nano* **12**, 1696-1703 (2018).

19. Muhammad, Z. *et al.* Transition from Semimetal to Semiconductor in $ZrTe_2$ Induced by Se Substitution. *ACS nano* **14**, 835-841 (2019).

20. Kar, I. *et al.* Metal-chalcogen bond-length induced electronic phase transition from semiconductor to topological semimetal in $ZrX_2$(X=Se and Te). *Phys. Rev. B* **101**, 165122 (2020).

21. Villaos, R. A. B. *et al.* Evolution of the Electronic Properties of $ZrX_2$(X = S, Se,

or Te) Thin Films under Varying Thickness. *J. Phys. Chem. C* **125**, 1134-1142 (2021).

22. Ren, M. Q. *et al.* Semiconductor-Metal Phase Transition and Emergent Charge Density Waves in 1T-ZrX$_2$(X = Se, Te) at the Two-Dimensional Limit. *Nano Lett* **22**, 476-484 (2022).

23. Yang, L.-N. *et al.* Coexistence of the charge density wave state and linearly dispersed energy band in 1T-ZrTe$_2$ monolayer. *Appl. Phys. Lett.* **120**, 073105 (2022).

24. Ugeda, M. M. *et al.* Giant bandgap renormalization and excitonic effects in a monolayer transition metal dichalcogenide semiconductor. *Nat. Mater.* **13**, 1091-1095 (2014).

25. Cercellier, H. *et al.* Evidence for an excitonic insulator phase in 1T-TiSe$_2$. *Phys. Rev. Lett.* **99**, 146403 (2007).

26. Monney, C. *et al.* Exciton Condensation Driving the Periodic Lattice Distortion of 1T-TiSe$_2$. *Phys. Rev. Lett.* **106**, 106404 (2011).

27. Hoesch, M. *et al.* Disorder Quenching of the Charge Density Wave in ZrTe$_3$. *Phys. Rev. Lett.* **122**, 017601 (2019).

28. Ye, Z. *et al.* Probing excitonic dark states in single-layer tungsten disulphide. *Nature* **513**, 214-218 (2014).

29. Chen, C., Singh, B., Lin, H. & Pereira, V. M. Reproduction of the Charge Density Wave Phase Diagram in 1T-TiSe$_2$ Exposes its Excitonic Character. *Phys. Rev. Lett.* **121**, 226602 (2018).

30. Zhang, K. N. N. *et al.* Widely tunable band gap in a multivalley semiconductor SnSe by potassium doping. *Phys. Rev. Mater.* **2**, 054603 (2018).

31. Rossnagel, K. On the origin of charge-density waves in select layered transition-metal dichalcogenides. *J. Phys.: Condens. Matter* **23**, 213001 (2011).

32. Jiang, Z. Y. *et al.* Spin-Triplet Excitonic Insulator: The Case of Semihydrogenated Graphene. *Phys. Rev. Lett.* **124** (2020).

33. Giannozzi, P. *et al.* Quantum ESPRESSO toward the exascale. *J. Chem. Phys.* **152**, 154105 (2020).

34. Mostofi, A. A. *et al.* An updated version of wannier90: A tool for obtaining maximally-localised Wannier functions. *Comput. Phys. Commun.* **185**, 2309-2310 (2014).

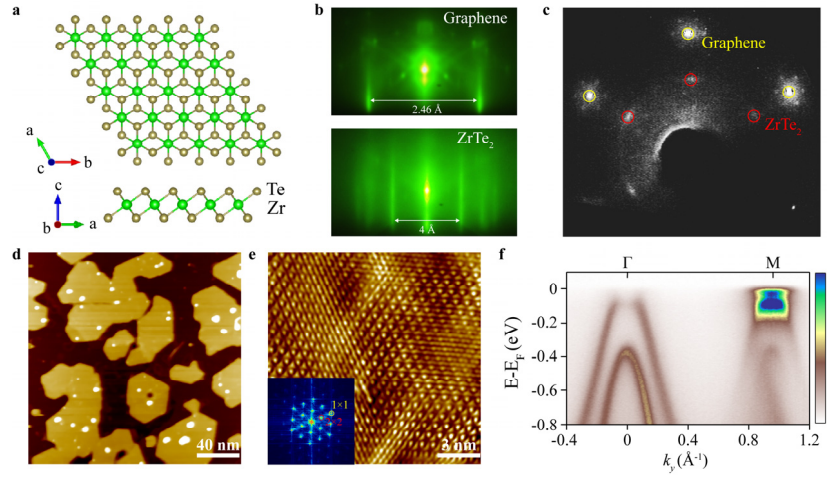

**Figure 1. Structure and characterization of epitaxially grown monolayer 1T-ZrTe₂.**
**a.** Schematic crystal structure of 1T-ZrTe₂. 1T-ZrTe₂ in the top (top panel) and side (bottom panel) views. **b.** RHEED patterns of graphene substrate (top panel) and sub-monolayer 1T-ZrTe₂ (bottom panel). **c.** LEED pattern after the film growing on the graphene. **d.** STM image of monolayer 1T-ZrTe₂ at 300 K [$V_{bias}$ = -2.8 V, $I$ = 200 pA, 200 nm×200 nm]. Dark regions correspond to the BLG substrate, and the 1T-ZrTe₂ layer is orange, which is in the form of islands with straight boundaries, dressed with a small number of particles that might be amorphous Zr-Te compounds or tiny 1T-ZrTe₂ islands. Some 1T-ZrTe₂ islands with a corner angle of 120° are identified, implying a hexagonal symmetry of 1T-ZrTe₂ single crystal. **e.** Atomically resolved STM image of a CDW-related 2×2 superstructure at 4.5 K [$V_{bias}$ = -50 mV, $I$ = 200 pA, 15 nm × 15nm]. The insert is the corresponding FFT. **f.** ARPES measured band structure along the Γ-M direction at $T$ = 15 K.

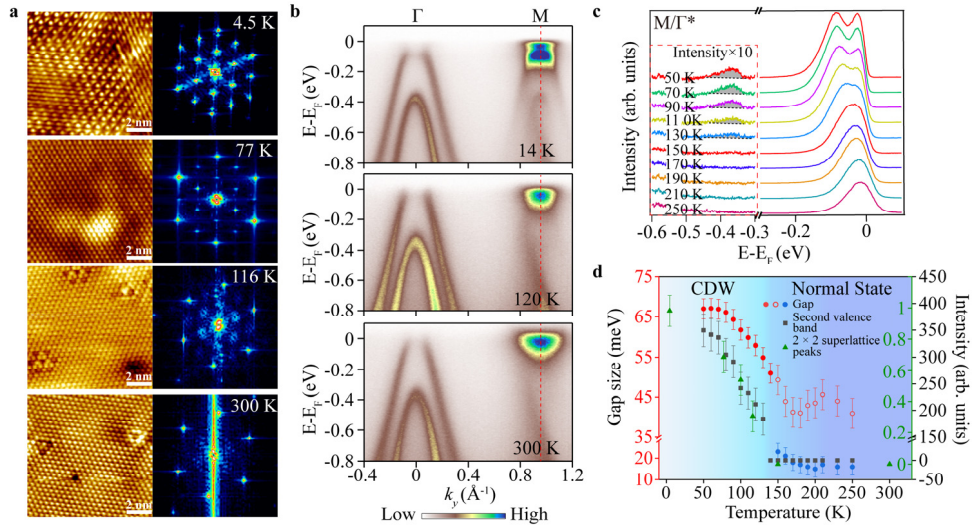

**Figure 2. Temperature dependence of CDW state in monolayer 1T-ZrTe$_2$. a.** Atomically resolved STM images of 10 nm × 10 nm and corresponding FFT images at 4.5 K [$V_{bias}$ = -50 mV, $I$ = 200 pA], 77 K [$V_{bias}$ = 50 mV, $I$ = 250 pA], 116 K [$V_{bias}$ = -150 mV, $I$ = 280 pA] and 300 K [$V_{bias}$ = -100 mV, $I$ = 300 pA]. **b.** The band structures measured along the experimental Γ-M direction at 14 K, 120 K, and 300 K. **c.** EDCs measured exactly at M/Γ* (correspond to red dashed lines in Fig. 2b), as a function of temperature. Three peaks are clearly recognized in the spectra measured at $T$ = 50 K, corresponding to one conduction band and two folded valence bands, respectively. **d.** The evolutions of the size of the bandgap, the intensity of the folded second valence band, and 2×2 superlattice peaks in STM with temperature. At the high-temperature region, it is hard to tell whether the folded first valence band merges into the conduction band or disappears. Different fitting strategies (circles with different colors, see the Supplementary Information for more detail) are employed, and the results show an identical trend.

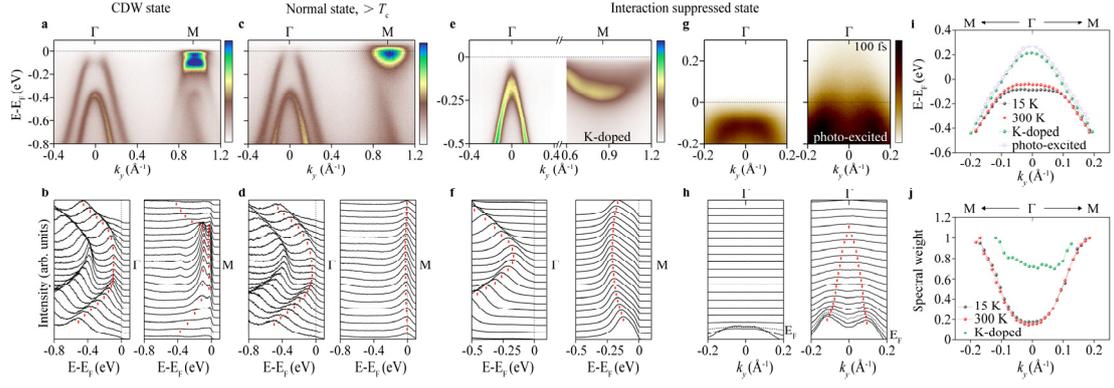

**Figure 3. The two-step process of the CDW formation in monolayer 1T-ZrTe$_2$. a. and c.** The band structures measured by ARPES along the Γ-M direction at 15 K (**a**) and 300 K (**c**) using 50.6 eV synchrotron radiation light. **e.** ARPES data along the Γ-M direction taken from surface K-doped sample. The corresponding EDCs are shown in **b**, **d,** and **f**, respectively. **g.** ARPES snapshots taken before optical pumping and at characteristic pump-probe delays at the Γ point. **h.** MDCs for the data shown in **g**. **i. and j.** Band position (bands are shifted by aligning the second valence band to 15K data for comparison) (**i**) and spectral weight (normalized by dividing by the maximum spectral weight) (**j**) of the first valence band at the Γ point along Γ-M direction. There is no apparent difference in the band dispersions and SW of the first valence band between the CDW (15 K) state and the normal state (300 K). In contrast, the interaction-suppressed state (photo-excited and K-doped) shows considerable band renormalization and SW recovery, respectively.

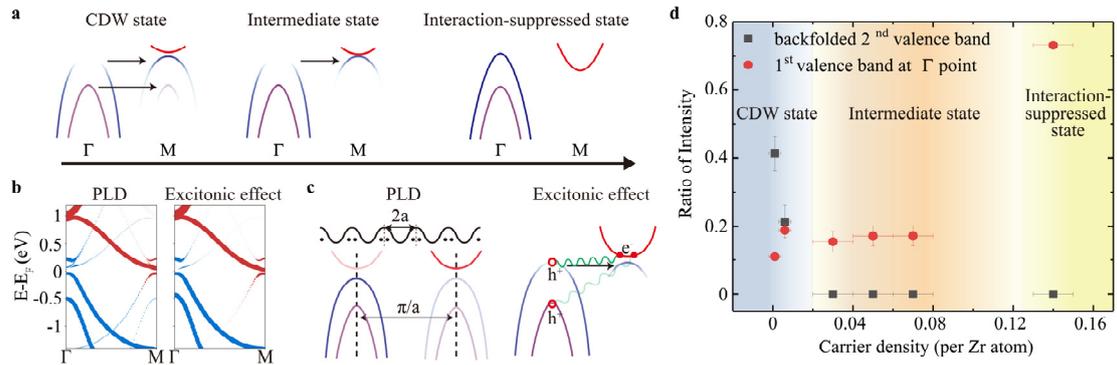

**Figure 4. The origin of CDW phase in monolayer 1T-ZrTe$_2$. a.** The schematic band diagram of the formation of the CDW state. **b.** The calculated band structure along the Γ-M direction for CDW driven by PLD and excitonic effect, respectively. **c.** The schematic diagram of PLD and excitonic effect, respectively. **d.** The evolution of electronic structure features of CDW state with doping amount.